\documentclass[preprint2]{aastex631}
\usepackage{natbib}
\usepackage{graphicx}
\usepackage{wasysym}
\usepackage{txfonts}
\usepackage{ulem}
\def\fH2{\mbox {f$_{{\rm H}_2}$}}

\def\EBV{\mbox{\rm{E(B-V)}}}

\def\vlsr{\mbox{${\rm v}_{\rm lsr}$}}

\def\nH2{\mbox{${\rm n}(\HH$)}}
\def\enH2{\mbox{$n_{(\HH$)}}}

\def\pccc{~{\rm cm}^{-3}} 
 
\def\pcc {\mbox{${~{\rm cm}^{-2}}$}}

\def\Tsub#1 {\mbox{${\rm T}_{\rm #1}$}}
\def\TK  {\Tsub K }
\def\TB  {\Tsub B }

\def\Tsp {\Tsub sp }

\def\degr{\mbox{$^{\rm o}$}}
\def\p{\mbox{$^+$}}

\def\h13cop{\mbox{{H$^{13}$CO\p}}}

\def\c3h2{\mbox{C$_3$H$_2$}}

 \def\R0{R$_0$}
\def\G0{\mbox{G$_0$}} 
  
  \def\{\mbox{deg{{}^\circ}}
\def\ddeg{{}^\circ\kern-.1em}

\def\kms{\mbox{km\,s$^{-1}$}}

\def\m1{\mbox{$^{-1}$}}

\def\E#1 {$10^{#1}$}
\def\E#1 {E{#1}}
\def\P#1,{$\nH2\TK~=~#1\times~10^4\pccc$~K}
\def\ec#1,#2,#3,{#1\,(#2)\E{#3}}

\def\H3{\mbox{H$_3$}}
\def\Lya{\mbox{Lyman-$\alpha$}}

\def\RH2{\mbox{R$_{\rm G}$}}
\def\g13{\mbox{g$_{13}$}}

\def\kHeH2{\mbox{$k_{ He-\HH}$}}

\def\tim#1,#2{\mbox{{$#1\times10^{#2}$}}}

\def\WH I{\mbox{$\Upsilon_{{\rm H I}}$}}

\def\L21{\mbox{{$\lambda$21cm}}}

\newcommand{\emm}[1]{\ensuremath{#1}}   
\newcommand{\emr}[1]{\emm{\mathrm{#1}}} 

\newcommand{\HH}{\emr{H_2}}

\shorttitle{High v$_{\rm lsr}$ means low \EBV/N(H I)}
\shortauthors{Harvey Liszt}

\begin{document}


\title{H I Kinematics and the E(B-V)/N(H I) ratio}


\author{Harvey Liszt}
\affil{National Radio Astronomy Observatory \\
        520 Edgemont Road, Charlottesville, VA 22903\\
        hliszt@nrao.edu}
\email{hliszt@nrao.edu}




\begin{abstract}

The $\lambda 21$cm H I emission that is used to trace the gas to dust ratio at
high Galactic latitudes has contributions from material beyond the Milky
Way disk, with uncertain and likely sub-Solar metallicity and dust content. 
These contributions can be isolated kinematically and their presence is clear 
for sightlines with small mean reddening $<$E(B-V)$>$ $\la$ 0.03 mag, which have mean ratios 
$<$N(H I)$>$/$<$E(B-V)$>$ that are 20-50\% above the high latitude Galactic 
average $<$N(H I)$>/<$E(B-V)$>=8.3\times10^{21}$cm$^{-2}$mag$^{-1}$. By mapping 
N(H I) and E(B-V) across H I High Velocity Cloud complexes and the Magellanic 
Clouds we show that the reddening of this kinematically-isolated gas is on 
average five times smaller per H I than the high latitude average. However, 
the aggregate contribution of this gas is small and
$<$N(H I)$>/<$E(B-V)$>=8.3\times10^{21}$cm$^{-2}$mag$^{-1}$ is the appropriate
value for Galactic gas seen at high latitude using the H I and reddening
measures employed here and in our previous work.

\end{abstract}



\section{Introduction}

The gas/dust ratio N(H)/\EBV\ is an important benchmark but its measured value 
is sensitive to context and technique. Two significantly different values are
derived from broad swaths of data respectively sampling gas in UV absorption 
toward early-type stars and gas seen in \L21\ H I emission from the diffuse 
interstellar medium in other directions.

When the atomic and molecular constituents 
of N(H) = N(H I)+2N(\HH) are measured in UV absorption toward early-type stars 
and compared with the associated photometric stellar reddening there is a 
tight relationship N(H)/\EBV\ $= 6.1-6.2\pm0.3\times10^{21}\pcc$ mag\m1\ 
\citep{LisGer23a,ShuPan24} very much like that first determined by \cite{BohSav+78},
$<$N(H)$>$/$<$\EBV$>$ $= 5.8\times10^{21}\pcc$ mag\m1\ using $Copernicus$.

The atomic gas/dust ratio N(H I)/\EBV\ should be an equivalent measure at 
small reddening \EBV\ $\la 0.08$ mag where the molecular gas fraction is 
negligible \citep{BohSav+78}, and indeed (Table 2 of \cite{LisGer23a}), 
$<$N(H I)$>$/$<$\EBV$>$ $= 6.2\times10^{21}\pcc$ mag\m1\ with N(H I) 
measured in \Lya\ absorption and compared with stellar photometry. But the 
scatter in individual stellar measurements is large owing to errors in the
photometric reddening, and the average merges a very small value $<$N(H I)$>$/$<$\EBV$>$ 
$= 4.1\times10^{21}\pcc$ mag\m1\ measured by the $Copernicus$ mission
with larger values $<$N(H I)$>$/$<$\EBV$>$ $\approx 6.9\times10^{21}\pcc$ mag\m1\
measured subsequently toward three times more stars by 
\cite{DipSav94} and \cite{GilShu+06}.

In any case, the above-quoted N(H)/\EBV\ and N(H I)/\EBV\ ratios measured toward 
stars are significantly smaller than the global low-reddening Galactic average 
$<$N(H I)$>$/$<$\EBV$>$ $= 8.3-8.8\times10^{21}\pcc$ mag\m1\ 
\citep{Lis14xEBV,Lis14yEBV,LenHen+17,ShuPan24} that is obtained when N(H I) is measured 
in \L21\ H I emission and compared with reddening equivalents derived at far infrared 
\citep{SchFin+98} or sub-mm \citep{Pla48} wavelengths. The difference 
may be related to the viewing geometries. The small scale heights of early-type 
stars constrain foreground absorbing material to the vicinity of the Galactic 
plane, where low reddening can occur along short sightlines or along
sightlines having a low mean density of intervening material (as was the 
case for $Copernicus$). 
By contrast, the low-reddening sightlines observed in \L21\ H I emission occur 
only at higher Galactic latitudes and preferentially sample material further 
from the plane.  As a result, the H I line profile integral at higher latitudes 
samples phenomena like high velocity clouds (HVC)\citep{Ver75} that do not occur 
near the Sun or in the Galactic disk.

This raises the question whether the elevated gas/dust ratios observed at
high Galactic latitudes arise through  destruction or stripping of dust in 
disk material \citep{DraSal78,SeaShu83,HenDra21} or from the 
presence of an admixture of low-reddening external gas. Addressing this question,
 \cite{ShuPan24} recently showed that inhomogeneities in the composition of 
the high latitude gas are manifested in variation of the gas/dust ratio 
with gas velocity.  They performed a hybrid measurement of N(\HH) in UV 
absorption toward QSO in combination with  N(H I) measured in \L21\ H I emission 
and \EBV\ taken from the FIR and sub-mm derived reddening equivalents cited above.  
In this way, they showed (their Table 7) that 9 low-reddening sightlines 
with relatively large contributions to N(H I) from gas at high negative 
velocity \vlsr\ $< -90$ \kms\ had elevated 
$<$N(H I)$>$/$<$\EBV$>$ $= 17 \times 10^{21}\pcc$ mag\m1\
in directions toward HVC Complex C \citep{Ver75}. They suggested that the 
HVC gas was material with low metallicity, 
small dust content and reddening. 

\begin{figure}
\includegraphics[height=12.2cm]{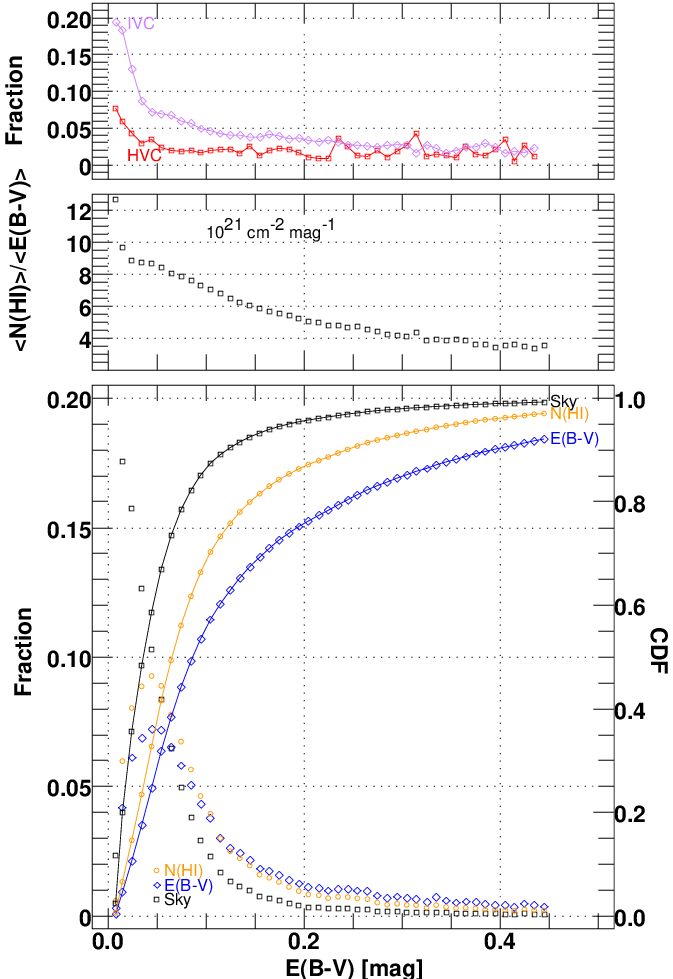}
 \caption{
Mean properties of the sky at $|b| > 20$\degr\ binned in 0.01 mag increments
of \EBV.  Top: Fraction of N(H I) in HVC at $|$\vlsr$|$ $>$ 90 \kms\ and IVC at
30 \kms\ $<$ $|$\vlsr$|$ $<$ 90 \kms.  Middle: $<$N(H I)$>$/$<$\EBV$>$. Bottom: 
distribution and cumulative distribution function (CDF) of the fraction of the 
sky, fraction of the reddening and fraction of the N(H I) binned in \EBV.}
\end{figure}

\begin{figure}
\includegraphics[height=8.5cm]{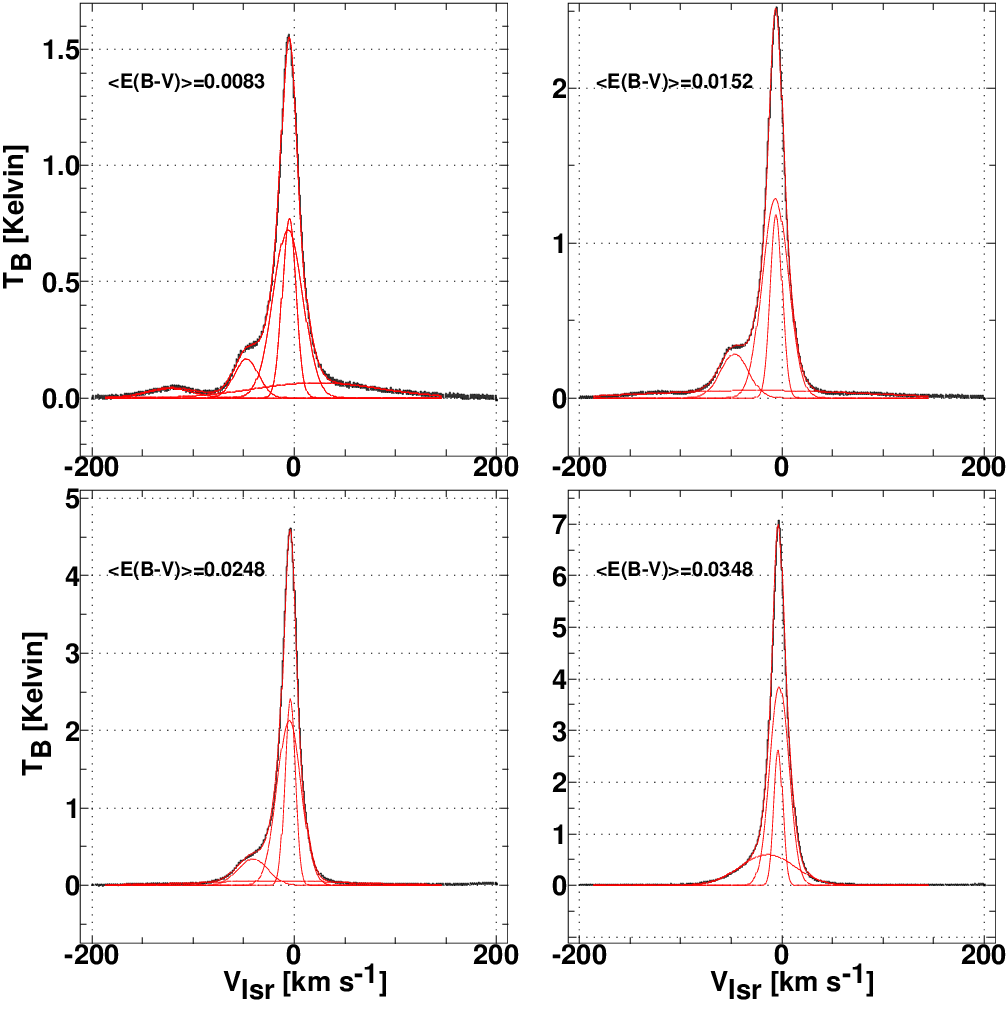}
 \caption{Mean \L21\ H I brightness profiles for the four lowest-extinction 
bins shown in Figure 1 with overlaid Gaussian decomposition.}
\end{figure}

\begin{figure*}
\includegraphics[height=18cm]{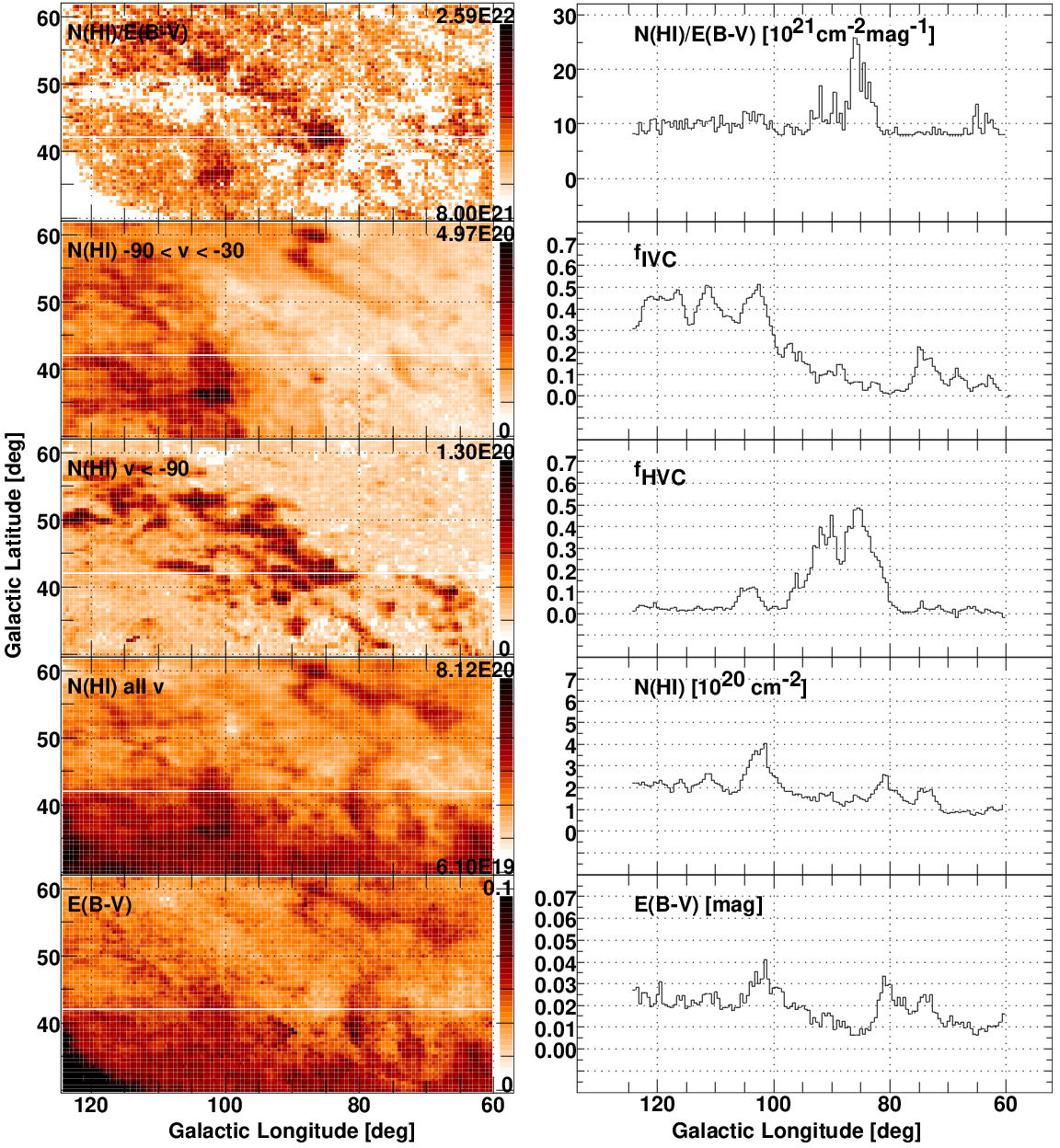}
\caption{
Properties of the sky over the region of HVC Complex C. Left top: N(H I)/\EBV\ scaled 
and clipped over the range $8 - 25 \times 10^{21} \pcc$ mag\m1. 
Left middle three panels: N(H I) integrated over velocity ranges indicated in 
each panel. Left bottom: \EBV\ scaled and clipped over the range 0 - 0.1 mag.
Right: Properties along  the indicated longitude strip at b=42\degr\ 
for the associated property mapped at left, where f$_{\rm HVC}$ is the fraction 
of the integrated H I profile at \vlsr\ $< -90$ \kms\ and f$_{\rm IVC}$ is the 
fraction in the velocity interval -90 $<$ \vlsr $<-30$ \kms.}
\end{figure*}

\begin{figure}
\includegraphics[height=9cm]{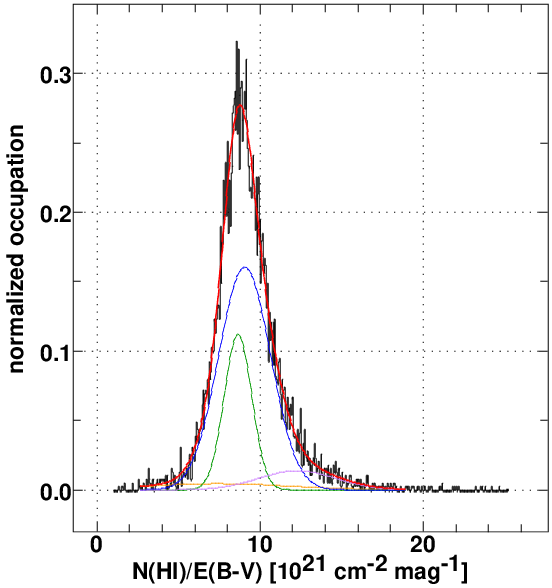}
  \caption{The distribution of reddening over the region shown in Figure 3,
 decomposed into several gaussian components.}
\end{figure}

\begin{figure*}
\includegraphics[height=7cm]{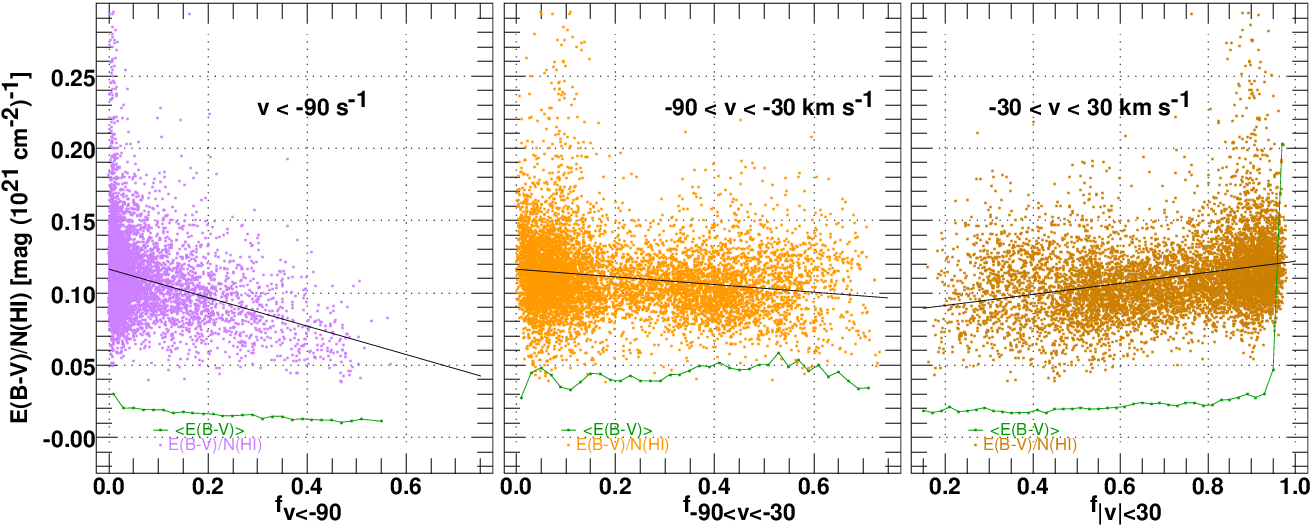}
\caption{
\EBV/N(H I) at individual pixels over the region of the HVC Complex C illustrated in 
Figure 3, plotted against the fraction of the total integrated brightness in three 
velocity ranges. The solid line in each panel is a least-squares regression fit and 
the green line is the mean reddening at each value of the fraction along the horizontal
axis.}
\end{figure*}

\begin{figure*}
\includegraphics[height=7cm]{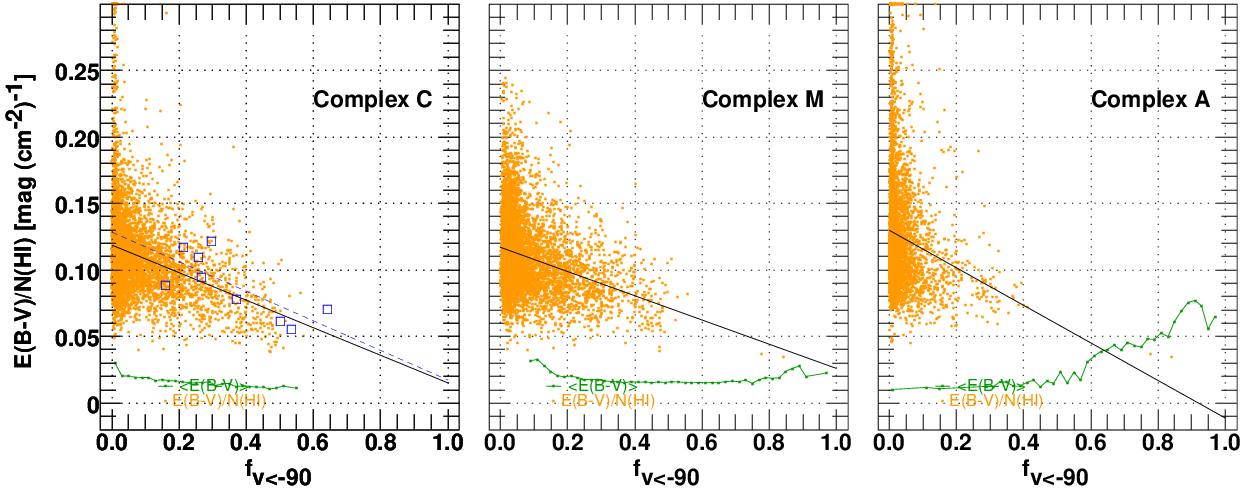}
\caption{
As in the left-most panel of Figure 5, but over an extended range of the H I fraction for
three HVC complexes. Blue rectangles and the blue dashed regression line in the left-most panel
represent the \EBV/N(H I) ratios derived using 353 GHz Planck dust opacity measurements
\citep{Pla48} by \cite{ShuPan24} (see their Table 7). 
}
\end{figure*}

\begin{figure*}
\includegraphics[height=8.5cm]{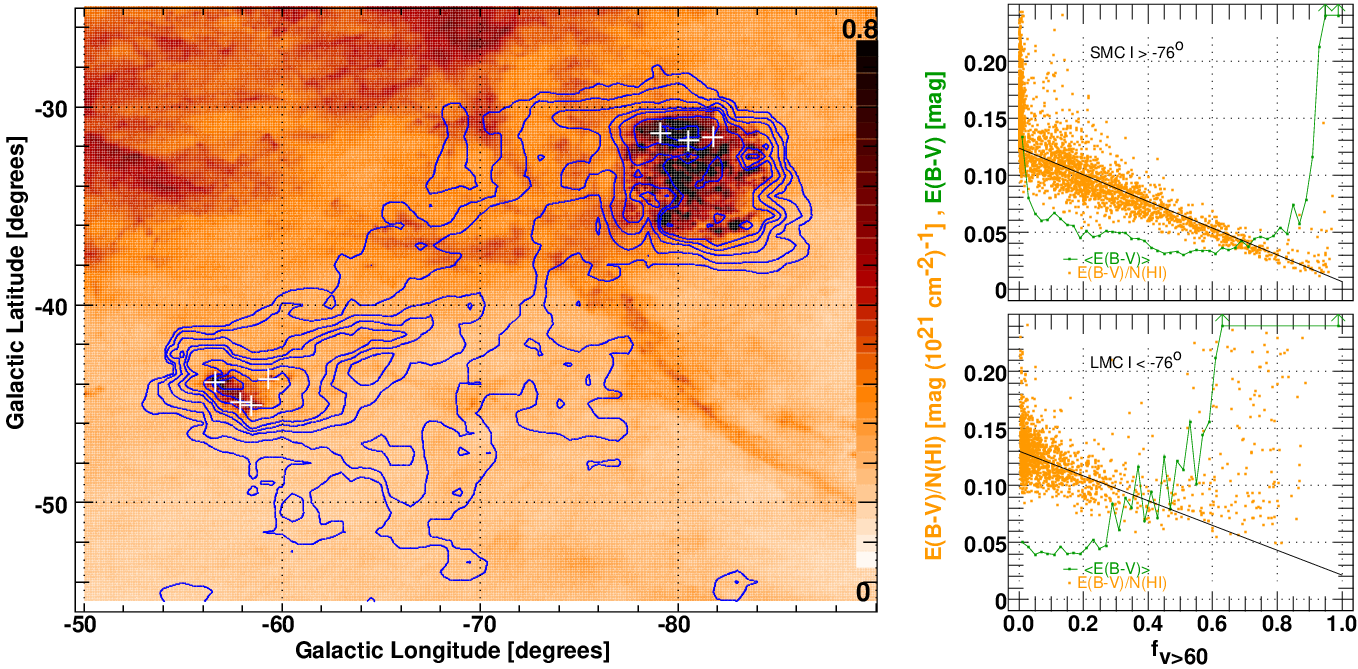}
\caption{\EBV, N(H I) and the dust/gas ratio toward the Magellanic 
Clouds.  Left: Contours of N(H I) integrated at \vlsr\ $\ge$ 60 \kms\ are shown 
at levels 1\%, 2\%, 4\%, 7\%, 10\%, 20\%, 40\% and 70\% of the peak 
N(H I) $ = 10^{22} \pcc$ and overlaid on a pseudocolor rendering of the reddening 
scaled between 0 and 0.8 mag.  The maximum reddenings in the SMC 
(the peak at $l \simeq -57$\degr) and LMC are 1.41 and 30.2 mag respectively 
and the mean reddening along the filament at middle-lower right is 
0.07$\pm0.03$ mag. White crosses denote regions of 
CO emission as noted in Section 3.1  Right: Plots 
of \EBV/N(H I) and $<$\EBV$>$ relative to the fraction of the integrated \L21\ 
H I brightness at v $>$ 60 \kms for the LMC (top) and SMC (bottom).  The solid 
line in each panel at right is a regression line calculated for the data 
at $<$\EBV$> \leq 0.1$ mag.  
\\
}
\end{figure*}

\begin{figure}
\includegraphics[height=6.8cm]{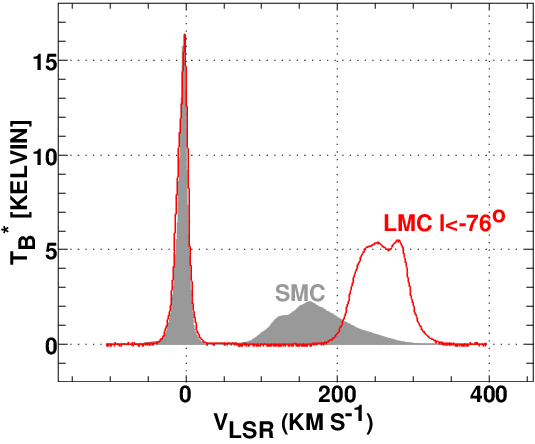}
\caption{
Mean spectra for the SMC  ($l \geq -76$\degr) and LMC ($l < -76$\degr) for
pixels in Figure 7 where N(H I) $\geq 10^{20}\pcc$.
}
\end{figure}

Such is the point of departure for the present work, which considers the extent
to which the high-latitude gas/dust ratio derived from \L21\ emission is affected 
by the inclusion of gas seen in emission outside the velocity range associated 
with Galactic rotation and material in the Galactic disk. Section 2 is a very 
brief summary of our data sources and methods. Section 3 is the actual discussion 
and Section 4 is a summary. 

\section{Data sources and conventions}

\subsection{\L21\ H I emission}

As the issue addressed here is the influence of low metallicity gas on our
previous determination of N(H I)/\EBV\ \citep{Lis14xEBV,Lis14yEBV}, we again use 
the Leiden/Argentine/Bonn (LAB) all-sky H I survey \citep{KalBur+05} with 0.6\degr\ 
angular resolution on a uniform 0.5\degr\ grid in Galactic coordinates. 

As before, H I column densities were derived assuming a spin temperature \Tsp\ = 
145 K.  The \L21\ optical depth $\tau$ was derived from the observed brightness 
temperature  
$$ \TB\ = \Tsp (1-\exp(-\tau)) $$
with \Tsp\ = 145 K and column densities N(H I) were computed from the fundamental 
physical relationship for the optical depth of the \L21\ line 
$$\tau = {\rm (dN(H I)/dv)}/(1.823\times10^{18}\pcc\Tsp)$$ 
with velocity in units of \kms\ \citep{Spi78}.

\subsection{\EBV}

Also as before, and for the sake of consistency, we use the FIR-derived reddening 
equivalents of \cite{SchFin+98}. As discussed by \cite{ShuPan24} and as pointed 
out somewhat more obscurely by \cite{Lis21}, the overall scale of the reddening 
derived from the best representation of the 353 GHz Planck dust opacity measurements 
\citep{Pla48} agrees within about 3\% with the work of \cite{SchFin+98} and not 
nearly so well with the later upward revision by a factor 1/0.86 recommended by 
\cite{SchFin11}.

\section{N(H I), E(B-V) and V$_{lsr}$}

Plots of N(H I) against \EBV\ in \cite{Lis14xEBV,Lis14yEBV} show a very tight 
linear relationship 
N(H I)/\EBV\ $= 8.3 \times 10^{21}\pcc$ mag\m1\ at 0.01 $\la$ \EBV\ $\la 0.08$ mag 
without an indication that N(H I) is being undersampled at the lowest \EBV\ due 
for instance to the presence of a noticeable proportion of ionized hydrogen. 

In such work, low-reddening sightlines  are only found at high Galactic latitudes 
and Figure 1 here explores the high latitude N(H I)-\EBV\ relationship in more 
detail. We averaged H I line profiles in 0.01 mag bins of 
\EBV\ at $|b| \geq 20 \degr$ and separated out the kinematic contributions of the 
so-called  high-velocity ($|$\vlsr$|$ $\geq 90$ \kms; HVC) and intermediate-velocity 
clouds (30 $\leq$ $|$\vlsr$|$ $\leq 90$ \kms; IVC) 
following \cite{Wak01}. The middle panel shows that 
sightlines in the bins with $<$\EBV$>$ $\leq 0.02$ mag have higher gas/dust ratios 
$<$N(H I)/\EBV$>$ $= 10-13 \times 10^{21}\pcc$ mag\m1\ and (upper panel) 
high proportions of gas at velocities characteristic of HVC.  

The  numerical values of the distributions plotted in Figure 1 are given in the 
Appendix.  The two lowest bins represent
fractions 0.023 and 0.176 of the sampled sky, fractions 0.006 and 0.060 of the total 
\EBV\ and fractions 0.003 and 0.042 of the aggregate N(H I). The HVC gas 
fraction asymptotically achieves a value of 2\% for $<$\EBV$>$ $> 0.05$ mag. 
The mean \L21\ H I emission profiles in the lowest 4 bins of \EBV\ 
are shown in Figure 2 where it is seen that the contribution of HVC gas is hard 
to distinguish for $<$\EBV$>$ $> 0.03$ mag where the emission around zero velocity 
is stronger.

The combined fractions of gas in HVC and IVC are 20-25\% in the lowest two bins but 
are too small overall to be responsible for the 40\% higher value of N(H I)/\EBV\ 
seen at high latitude (ie the multipliers 8.3 vs 6.2). This implies that stripping
or destruction of dust in Galactic disk gas at large z-heights is responsible, but the 
kinematic information in the H I profile nevertheless contains information that can 
be exploited to refine the appropriate high-latitude average gas/reddening ratios 
for the disk and high-velocity  gas.

Figure 3 at left shows properties of the sky over a large region containing HVC 
Complex C (see also \cite{ShuPan24}). The HVC gas column density in the middle panel 
is clearly associated 
with an elongated region of high N(H I)/\EBV\ at top that is not apparent in the 
maps of \EBV\ and integrated N(H I) at bottom.  A small region of high N(H I)/\EBV\ 
in the top left near (l,b) = (102\degr,37\degr) is associated with a region of high 
IVC gas column density. Overall the IVC gas is not strongly deficient in \EBV\
and the velocity boundary between HVC and IVC gas is only loosely defined.   

Figure 3 at right shows small-scale behaviour along longitude strips at 
b=42\degr\ for each of the properties mapped at left, replacing the HVC and
IVC column densities by the fractions of the integrated H I line profile (H I 
column density) that reside in the HVC and IVC components. The coincident minimum 
in \EBV\ and maximum in the HVC fraction at $l = 85$\degr\ combine to produce
a narrow region of very high N(H I)/\EBV. The strongly-varying IVC fraction has 
little effect on the N(H I)/\EBV\ ratio, suggesting that its metallicity and 
dust/gas content are similar to the Galactic average at these latitudes.

The overall character of the reddening in the region of HVC Complex C is shown in 
Figure 4 where the distribution of values of N(H I)/\EBV\ is decomposed into 
gaussian components.  In this view the bulk of the gas has 
N(H I)/\EBV\ $= 8-9 \times 10^{21}\pcc$ mag\m1\ with a small admixture in a 
broad component centered at $12 \times 10^{21}\pcc$ mag\m1.  Smaller values of 
N(H I)/\EBV\ are associated with the higher-reddening region at the lower left 
corner of map nearer the Galactic plane.

The variation of N(H I)/\EBV\ over the region of high HVC gas fraction at l=100 - 115\degr\ 
in Figure 3 at right somewhat complicates a simple picture where the N(H I)/\EBV\
ratio is determined by a weighted average of disk gas with high metallicity
and dust content and an HVC component with low  metallicity and near-zero reddening. 
Nonetheless, it does appear that the HVC gas on average contributes very little 
reddening.  Shown in Figure 5 are plots of the reddening per atom \EBV/N(H I) (both
total, N(H I) is the entire profile integral in each case) 
as it varies with the fractions of the integrated emission or column density
in the HVC, IVC and disk ($|$\vlsr$|$ $\leq 30$ \kms) gas components.  

The mean reddening per H I declines sharply with the HVC gas fraction and only weakly 
with the fraction of IVC gas that on average must have dust content that is not greatly 
different from ``normal''. The reddening per atom varies in the opposite sense with 
the fraction of gas around 0-velocity, in compensatory fashion. The regression lines 
intercept the vertical axis near the equivalents of the canonical high Galactic latitude 
value N(H I)/\EBV\ $\approx 8.3 \times 10^{21}\pcc$ mag\m1\ for 0 gas fractions in
the HVC and IVC components or 100\% of the low-velocity (see Table 1 as discussed below).

Figure 6 likewise plots the \EBV/N(H I) ratio against the HVC gas fraction across 
the three HVC complexes C, M, and A, extended 
over the full range of the horizontal axis.  The behavior is similar in all cases,
indicating on average very little contribution to \EBV\ from gas at v $\leq -90$ \kms\
as inferred from the small values of \EBV/N(H I) extrapolated to unit HVC gas fraction.
This is quantified in Table 1 as discussed in Section 3.2 with the inclusion of gas seen 
around the Magellanic Clouds. 
This is particularly interesting for the region containing HVC Complex M where 
\cite{SchVer22} and \cite{VerSch23} have presented a convincing case that the gas 
kinematics are related to a supernova remnant destroying dust in disk 
gas that otherwise would be expected to have a normal gas/dust ratio.

\subsection{The Magellanic Clouds}

The region of the sky  around the Large and Small Magellanic Clouds (LMC and SMC; 
Figure 7) presents examples 
where the gas velocity is at least somewhat associated with identifiable systems 
whose metallicities have been measured, so the effects of differing metallicity 
may be explored within the Clouds and in comparision with Milky Way material 
having Solar metallicity observed in the same directions. Typical metallicities
are one-half and one-fifth Solar for the Large and Small Clouds, respectively.

Shown in Figure 7 at left is a map of the H I column density at \vlsr\ $>$ 60 \kms\
where the stellar body of the SMC overlays the peak at $l \simeq -57$\degr and 
we have indicated regions where \cite{SalRub+23}
mapped CO emission around two positions in the Bar and in N83 and NGC602. \L21\ 
H I emission spectra in this portion of the SMC are among the brightest in the entire 
sky, with a peak T$_{\rm B} \approx 125$ K, presumably because the small metallicity 
and low dust content of the SMC inhibit the conversion of atomic to molecular 
hydrogen even with very large gas columns.  The LMC overlays the H I and \EBV\  
peaks at $l \la -78$\degr\ and the crosses there represent regions where 
\cite{FinMol+22} observed CO emission in the Ridge, 30Dor and N159.  Mean spectra
illustrating the H I kinematics of both Clouds are shown in Figure 8. 

Figure 7 at right shows strong differences in the H I, reddening and dust/gas ratios 
between the regions on either side of $l = -76$\degr. The region at 
$l > -76$\degr\ including the SMC  at top right has only very limited regions 
where \EBV\ $> 0.05$ mag and the variation of the \EBV/N(H I) ratio 
resembles those shown for the HVC Complexes in Figures 6 and 7 while
extending to fractional occupations near unity having very small \EBV/N(H I).
By contrast, much of the surface area of the H I distribution in the LMC
at $l < -76$\degr\  has
appreciable reddening \EBV\ $>$ 0.1 mag where the dust/gas ratio 
\EBV/N(H I) is at least 2/3 that of the Galactic gas seen in this portion of the sky.
The reddening in regions of very high apparent \EBV\ toward the LMC have been 
distorted by FIR emission associated with the regions of strong star formation
in the LMC. To remove the distortions introduced by star formation,
regression fits were limited to pixels with \EBV\ $<$ 0.1 mag.

\subsection{Limiting values}

Shown in Table 1 are the limiting values for the gas/dust and dust/gas ratios
in the HVC complexes and Magellanic Clouds, derived from the regression lines
shown in Figures 6 and 7. The limiting values for unit spectrum occupancy of 
HVC and Magellanic Cloud gas are typically 5-7 times smaller with a weighted
mean difference of a factor $5.3\pm1.7$.  The values extrapolated to 0 spectrum 
occupation by high-velocity gas cluster around the old value and have a weighted mean 
N(H I)/\EBV\ $= 8.3\pm0.4 \times 10^{21}\pcc$ mag\m1, indicating that the
old value was not materially influenced by the presence of low metallicity
material.

\subsection{Is \EBV\ reliably estimated in HVC gas?}
 The reddening per H-atom assigned to HVC gas by \cite{SchFin+98} is some five times
smaller than for Galactic disk gas (Table 1) and a similar  result is found when the
reddening is derived from Planck dust opacity measurements \citep{Pla48}.  Shown
at left in Figure 6 and summarized in  Table 1 are the \EBV/N(H I) values derived
toward HVC Complex C by \citep{ShuPan24} using Planck data. The nominal gradient is
slightly steeper but differences in the limiting values are not statistically significant.  
We conclude that the small reddening per H I in HVC gas is a real physical effect.

\section{Summary}

The column densities of neutral atomic and molecular hydrogen N(H I) and N(\HH) 
were measured in UV absorption toward 80 early-type stars 
using the $Copernicus$ satellite and a tight relationship was found between 
N(H) = N(H I)+2N(\HH) and the photometric reddenning,  
$<$N(H)$>/<$\EBV$> = 5.8 \times 10^{21}\pcc$ mag\m1. 
This ratio of gas column density to dust reddening became an important Milky Way 
 benchmark at Solar metallicity, with many practical applications owing to the 
relative ease with which \EBV\ can be measured or derived, compared to either or 
both N(H I) and N(\HH).

The ratio N(H I)/\EBV\ would be expected to have the same value at low reddening 
\EBV\ $\la 0.08$ mag where the \HH\ fraction is negligible, but a confirmatory
determination of the ratio toward stars is complicated by large scatter in the 
phototometric reddening and an independent determination using N(H I) derived 
from \L21\  emission and sub-mm/FIR-derived \EBV\ equivalents found a noticeably 
different value $<$N(H I)$>/<$\EBV$> = 8.3 \times 10^{21}\pcc$ mag\m1\ at 
Galactic latitudes above $|b| = 20$\degr\ where sightlines with such low
reddening are found.

Observations of \L21\ H I emission at high Galactic latitude sample gas at large 
z-height in the disk and gas outside the Galactic disk that is not present along 
sightlines toward early-type stars that, as a population, have small Galactic scale 
heights. The gas/reddening ratio determined over long path lengths at high latitude 
could be affected by including disk gas stripped of dust at large heights in the Milky 
Way disk or by the inclusion of low-metallicity gas having an intrinsically small dust 
content outside the disk.   

Disk gas and gas external to the disk can be distinguished in H I kinematics. In 
Figures 1 and 2 we showed that there is a higher proportion of high and intermediate 
velocity H I along sightlines at \EBV\ $\le 0.03$ mag having higher N(H I)/\EBV\ 
ratios  (see Table 2). However, the fraction of kinematically-distinguishable 
gas is too small to explain the 40\% larger gas/reddening ratio seen in \L21\ emission 
at high latitudes. 

That being said, a reddening deficit in high speed gas can manifest itself quite strongly 
on small scales locally (Figure 3) when the H I gas fraction is large at velocities 
$|$\vlsr$|$ $>$ 90 \kms.  The so-called intermediate velocity H I at 
30 $\la$ $|$\vlsr$|$ $\la$ 90 \kms\ does not have a strong reddening deficit (Figure 5)
and Galactic gas at $|$\vlsr$|$ $\ga$ 30 \kms\ can appear at latitudes 
$|b|$ $\la 30$\degr\ through normal Galactic rotation. 

We determined the reddening deficit in high speed gas by correlating the reddening 
per H I \EBV/N(H I) with the high speed gas fraction across H I High Velocity
Cloud Xomplexes C, M, and A in Figure 6 and over the region of the Magellanic 
Clouds and the Magellanic Bridge in Figure 7.  As summarized numerically in
Table 1 the results are consistent in showing a deficit of a factor $5.3\pm1.7$
in \EBV/N(H I) and in confirming the multiplier 8.3 in the previously determined 
relationship $<$N(H I)$>/<$\EBV$> = 8.3 \times 10^{21}\pcc$ mag\m1\
using the measures of N(H I) and \EBV\ employed in this work.

\begin{table*}
\caption{Limiting values of \EBV/N(H I) and N(H I)/\EBV$^1$  }
{
\small
\begin{tabular}{lccc}
\hline
Source &\EBV/N(H I) &\EBV/N(H I) & N(H I)/\EBV\\
Limit$^2$  & f = 0 &  f = 1 & f = 0 \\
\hline
Complex C & 0.1186(0.00052) & 0.0150(0.0051) & 8.43 \\
Complex M & 0.1174(0.00036) & 0.0263(0.0033)& 8.52  \\
Complex A & 0.1302(0.00084) & -0.0110(0.0140) & 7.68  \\
SMC/Bridge  & 0.1170(0.00073) & 0.0066(0.0020)& 8.54 \\
LMC        & 0.1303(0.00117) & 0.0215(0.0067)& 7.67  \\
\hline
Weighted mean& 0.1198(0.00470) & 0.0222(0.0072) & 8.34  \\
\hline 
Complex C \citep{ShuPan24} &0.1283(0.0144) & 0.0172 (0.0360) & 7.79\\
\hline    
\end{tabular}
\\
$^1$ N(H I) is expressed in units of $10^{21}$ H I $\pcc$ and \EBV\ in mag\\
$^2$ fractions f of N(H I) at v $< -90$ \kms or v $> 60$ \kms\ for HVC and Magellanic Clouds, respectively
}
\end{table*}

\begin{acknowledgments}
  The National Radio Astronomy Observatory is a facility of the National
  Science Foundation operated under contract by Associated Universities, Inc.
  I thank the referee for helpful comments that clarified some important points.
 
\end{acknowledgments}

\appendix

\section{Numerical values}

Numerical values for the data plotted in Figure 1 are given in Table 2.
f$_{\rm HVC}$ and f$_{\rm IVC}$ are the fractions of the mean H I profile occupied by 
HVC and IVC gas at $|$\vlsr$|$ $>$ 90 \kms\
and 30 \kms\ $<$ $|$\vlsr $|$ $< 90$ \kms, respectively. 
f(sky), f(\EBV) and f(N(H I))
are the fractions of the sky, aggregate reddening and aggregate H I column density in the reddening
bin represented in each row and the cdf are the respectrum cumulative sum of these fractions.

\begin{table*}
\caption{Numerical values for the quantities plotted in Figure 1}
{
\tiny
\begin{tabular}{ccccccccccc}
\hline
\hline
\EBV &$|b|$ &N(H I)/\EBV&{f$_{\rm HVC}$}$^1$&{f$_{\rm IVC}$}$^1$&f(sky)$^2$&f(\EBV)$^2$&f(H I)$^2$&cdf(sky)&cdf(\EBV)&cdf(N(H I))\\
mag&\degr&$10^{21}\pcc$mag \m1&&&&&&&\\
\hline 
0.0083& 57.94& 12.654& 0.077& 0.194& 0.0234& 0.0057& 0.0030& 0.0234&0.0057&0.0030\\
0.0152& 61.35& 9.6591& 0.058& 0.183& 0.1757& 0.0597& 0.0418& 0.1991&0.0654&0.0448\\
0.0248& 55.82& 8.8715& 0.043& 0.130& 0.1575& 0.0802& 0.0611& 0.3566&0.1457&0.1059\\
0.0348& 46.72& 8.7065& 0.029& 0.086& 0.1266& 0.0886& 0.0688& 0.4832&0.2343&0.1747\\
0.0448& 40.35& 8.6869& 0.034& 0.072& 0.1029& 0.0926& 0.0720& 0.5861&0.3269&0.2467\\
0.0548& 36.01& 8.3869& 0.023& 0.069& 0.0836& 0.0888& 0.0716& 0.6697&0.4158&0.3182\\
0.0648& 34.02& 8.0507& 0.019& 0.067& 0.0645& 0.0777& 0.0652& 0.7341&0.4935&0.3834\\
0.0748& 33.28& 7.8287& 0.018& 0.059& 0.0497& 0.0672& 0.0580& 0.7838&0.5606&0.4414\\
0.0848& 32.91& 7.5879& 0.019& 0.056& 0.0380& 0.0565& 0.0503& 0.8218&0.6172&0.4917\\
0.0948& 32.10& 7.2661& 0.016& 0.049& 0.0291& 0.0462& 0.0430& 0.8509&0.6634&0.5347\\
0.1047& 31.52& 7.0532& 0.019& 0.046& 0.0231& 0.0394& 0.0377& 0.8740&0.7028&0.5724\\
0.1148& 30.66& 6.7653& 0.020& 0.042& 0.0167& 0.0299& 0.0299& 0.8907&0.7327&0.6023\\
0.1250& 29.95& 6.4681& 0.021& 0.040& 0.0134& 0.0249& 0.0260& 0.9040&0.7576&0.6283\\
0.1348& 30.09& 6.2215& 0.015& 0.040& 0.0115& 0.0222& 0.0241& 0.9155&0.7799&0.6525\\
0.1448& 29.75& 6.0510& 0.025& 0.037& 0.0096& 0.0194& 0.0216& 0.9251&0.7993&0.6741\\
0.1549& 29.24& 5.8634& 0.012& 0.037& 0.0075& 0.0158& 0.0182& 0.9326&0.8150&0.6923\\
0.1649& 29.06& 5.6698& 0.019& 0.041& 0.0067& 0.0146& 0.0173& 0.9393&0.8296&0.7096\\
0.1746& 28.19& 5.5565& 0.022& 0.039& 0.0058& 0.0130& 0.0158& 0.9451&0.8426&0.7254\\
0.1849& 28.36& 5.4019& 0.021& 0.035& 0.0048& 0.0111& 0.0138& 0.9499&0.8536&0.7392\\
0.1949& 28.86& 5.2126& 0.016& 0.036& 0.0041& 0.0095& 0.0124& 0.9540&0.8632&0.7516\\
0.2047& 28.49& 5.0432& 0.010& 0.033& 0.0034& 0.0082& 0.0110& 0.9574&0.8714&0.7626\\
0.2152& 29.08& 4.9869& 0.008& 0.031& 0.0032& 0.0079& 0.0107& 0.9606&0.8793&0.7733\\
0.2248& 28.89& 4.7794& 0.008& 0.033& 0.0028& 0.0068& 0.0097& 0.9634&0.8861&0.7830\\
0.2350& 29.23& 4.7664& 0.036& 0.030& 0.0028& 0.0073& 0.0104& 0.9662&0.8934&0.7933\\
0.2448& 28.59& 4.6846& 0.024& 0.027& 0.0027& 0.0072& 0.0104& 0.9689&0.9007&0.8038\\
0.2546& 28.89& 4.7165& 0.012& 0.026& 0.0024& 0.0067& 0.0097& 0.9714&0.9074&0.8134\\
0.2649& 28.78& 4.5209& 0.011& 0.025& 0.0023& 0.0064& 0.0096& 0.9737&0.9138&0.8230\\
0.2749& 28.56& 4.4069& 0.019& 0.024& 0.0018& 0.0051& 0.0078& 0.9755&0.9189&0.8308\\
0.2849& 27.84& 4.2063& 0.010& 0.026& 0.0016& 0.0043& 0.0069& 0.9771&0.9233&0.8378\\
0.2950& 28.14& 4.1864& 0.018& 0.027& 0.0016& 0.0046& 0.0074& 0.9787&0.9278&0.8451\\
0.3050& 28.80& 4.1222& 0.026& 0.027& 0.0014& 0.0042& 0.0068& 0.9801&0.9320&0.8520\\
0.3145& 28.17& 4.3646& 0.042& 0.016& 0.0013& 0.0041& 0.0064& 0.9814&0.9361&0.8584\\
0.3250& 27.17& 3.8315& 0.011& 0.027& 0.0011& 0.0031& 0.0054& 0.9825&0.9392&0.8638\\
0.3348& 28.51& 3.9258& 0.014& 0.022& 0.0014& 0.0041& 0.0071& 0.9838&0.9433&0.8709\\
0.3450& 28.22& 3.8786& 0.012& 0.016& 0.0011& 0.0034& 0.0059& 0.9849&0.9467&0.8767\\
0.3551& 27.08& 3.8915& 0.009& 0.019& 0.0009& 0.0029& 0.0051& 0.9858&0.9496&0.8818\\
0.3649& 27.58& 3.8488& 0.025& 0.024& 0.0010& 0.0031& 0.0055& 0.9868&0.9527&0.8873\\
0.3751& 26.66& 3.5868& 0.014& 0.024& 0.0009& 0.0028& 0.0052& 0.9877&0.9555&0.8925\\
0.3847& 28.21& 3.5819& 0.012& 0.029& 0.0007& 0.0023& 0.0044& 0.9884&0.9578&0.8969\\
0.3949& 27.03& 3.4079& 0.020& 0.024& 0.0007& 0.0022& 0.0043& 0.9891&0.9600&0.9012\\
0.4052& 28.41& 3.5324& 0.034& 0.015& 0.0006& 0.0021& 0.0041& 0.9898&0.9621&0.9053\\
0.4146& 26.98& 3.5868& 0.005& 0.018& 0.0005& 0.0018& 0.0034& 0.9903&0.9639&0.9086\\
0.4249& 27.51& 3.4885& 0.026& 0.016& 0.0007& 0.0024& 0.0047& 0.9910&0.9664&0.9134\\
0.4347& 26.01& 3.3457& 0.011& 0.023& 0.0006& 0.0020& 0.0041& 0.9916&0.9684&0.9175\\
0.4447& 27.30& 3.5281& 0.075& 0.017& 0.0005& 0.0019& 0.0036& 0.9921&0.9702&0.9210\\
0.6423& 25.10& 2.5448& 0.049& 0.016& 0.0079& 0.0298& 0.0790& 1.0000&1.0000&1.0000\\
\hline
\end{tabular}
\\
$^1$ f$_{\rm HVC}$ and f$_{\rm IVC}$ are the fraction of H I in HVC and IVC in each bin\\
$^2$ f(sky), f(\EBV) and f(H I) are the fractions of sky, \EBV\ and N(H I) in each bin \\
\\
}
\end{table*}


\bibliographystyle{apj}

\end{document}